\documentclass[prl,aps,twocolumn,showpacs,preprintnumbers,amsmath,amssymb]{revtex4}

\usepackage{graphicx}
\usepackage{dcolumn}
\usepackage{color}

\begin{document}

\def\EF{$E_\textrm{F}$}
\def\cred{\color{red}}
\def\cblue{\color{blue}}
\definecolor{dkgreen}{rgb}{0.31,0.49,0.16}
\def\cgreen{\color{dkgreen}}

\title{Multiple topological Dirac cones in a mixed-valent
Kondo semimetal: $g$-SmS}

\author{Chang-Jong Kang$^1$}
\email[Co-first authors]{}
\author{Dong-Choon Ryu$^2$}
\email[Co-first authors]{}
\author{Junwon Kim$^2$}
\author{Kyoo Kim$^{2,3}$}
\author{J.-S. Kang$^4$}
\author{J. D. Denlinger$^5$}
\author{G. Kotliar$^{1,6}$}
\author{B. I. Min$^2$}
\email[Corresponding author: ]{bimin@postech.ac.kr}

\affiliation{
$^1$Department of Physics and Astronomy, Rutgers University,
    Piscataway, New Jersey 08854, USA \\
$^2$Department of Physics, Pohang University of Science and Technology,
    Pohang 37673, Korea \\
$^3$MPPHC$\_$CPM, Pohang University of Science and Technology,
	Pohang 37673, Korea \\
$^4$Department of Physics, The Catholic University of Korea,
	Bucheon 14662, Korea\\
$^5$Advanced Light Source, Lawrence Berkeley Laboratory,
	Berkeley, California 94720, USA \\
$^6$Condensed Matter Physics and Materials Science Department,
    Brookhaven National Laboratory, Upton, New York 11973, USA
}
\date{\today}

\begin{abstract}
We demonstrate theoretically that
the golden phase of SmS ($g$-SmS),
a correlated mixed-valent system,
exhibits nontrivial surface states with diverse topology.
It turns out that this material is an ideal playground to
investigate different band topologies in different surface terminations.
We have explored surface states on three different
(001), (111), and (110) surface terminations.
Topological signature in the (001) surface is not apparent
due to a hidden Dirac cone inside the bulk-projected bands.
In contrast, the (111) surface shows a clear
gapless Dirac cone in the gap region,
demonstrating the unambiguous topological Kondo nature of $g$-SmS.
Most interestingly, the (110) surface exhibits both
topological-insulator-type and
topological-crystalline-insulator (TCI)-type
surface states simultaneously.
Two different types of double Dirac cones, Rashba-type and TCI-type,
realized on the (001) and (110) surfaces, respectively,
are analyzed with the mirror eigenvalues and mirror Chern numbers
obtained from the model-independent \emph{ab initio} band calculations.
\end{abstract}

\maketitle


\emph{Introduction}.
Topological insulators have been studied intensively in recent years
as a new phase of quantum matter and
for possible applications to spintronics
and quantum computing \cite{Hasan10,Qi11}.
The physical significance of topological insulators is that
they possess metallic surface states that are protected by time-reversal symmetry,
which leads to the exhibition of robust surface states
under any perturbations without breaking the symmetry \cite{Hasan10,Qi11}.
Since the theoretical study proposed that nontrivial topology emerges
in strongly correlated mixed-valent or Kondo systems \cite{Dzero10,Dzero12},
subsequent experiments have been conducted
to examine the topological properties
in a candidate material SmB$_6$
\cite{Wolgast13,DJKim13,DJKim14,Robler13,Jiao16,
Li14science,Xu13,Neupane13,Denlinger13,Jiang13,
Zhu13,Min14,Xu14,Hlawenka18}.
Several angle-resolved photoemission spectroscopy (ARPES) measurements
confirmed a topologically-driven metallic surface state
\cite{Denlinger13,Min14,Xu13,Jiang13,Neupane13}
and its spin helicity in the (001) surface \cite{Xu14},
where these topological properties are induced solely
from a single Dirac cone.
But a few ARPES reports claim that
the observed surface states are just
trivial \cite{Zhu13,Hlawenka18}.
So the controversy still remains.

\begin{figure}[b]
\includegraphics[width=7.5 cm]{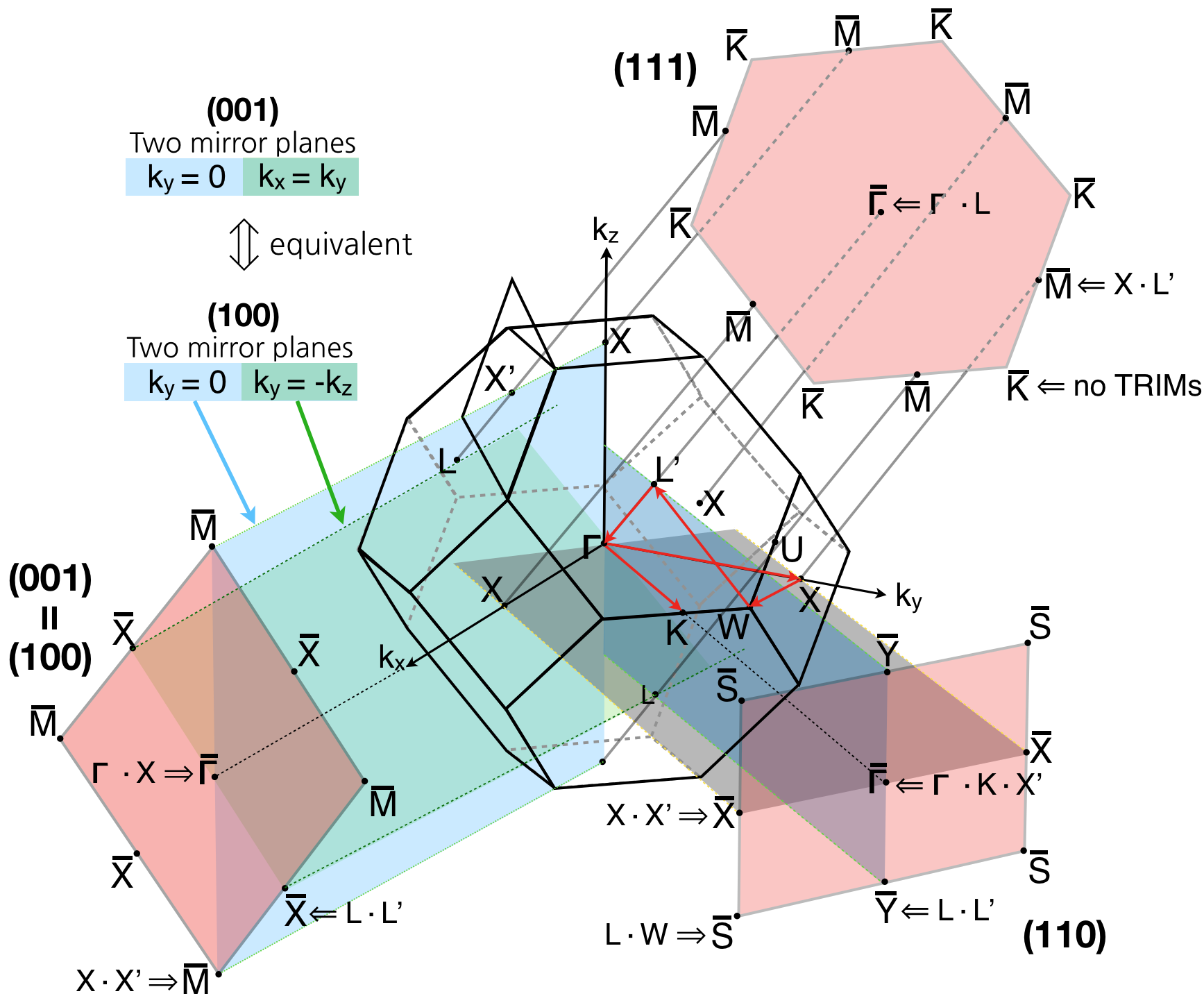}
\caption{(Color Online)
Bulk and surface BZs of fcc $g$-SmS.
There are mirror-symmetry lines
along $\bar{\Gamma}-\bar{M}$ and $\bar{\Gamma}-\bar{X}$
on the (100) surface BZ,
and along  $\bar{\Gamma}-\bar{X}$ and $\bar{\Gamma}-\bar{Y}$
on the (110) surface BZ.
}
\label{BZ}
\end{figure}

\begin{figure*}[t]
\includegraphics[width=17 cm]{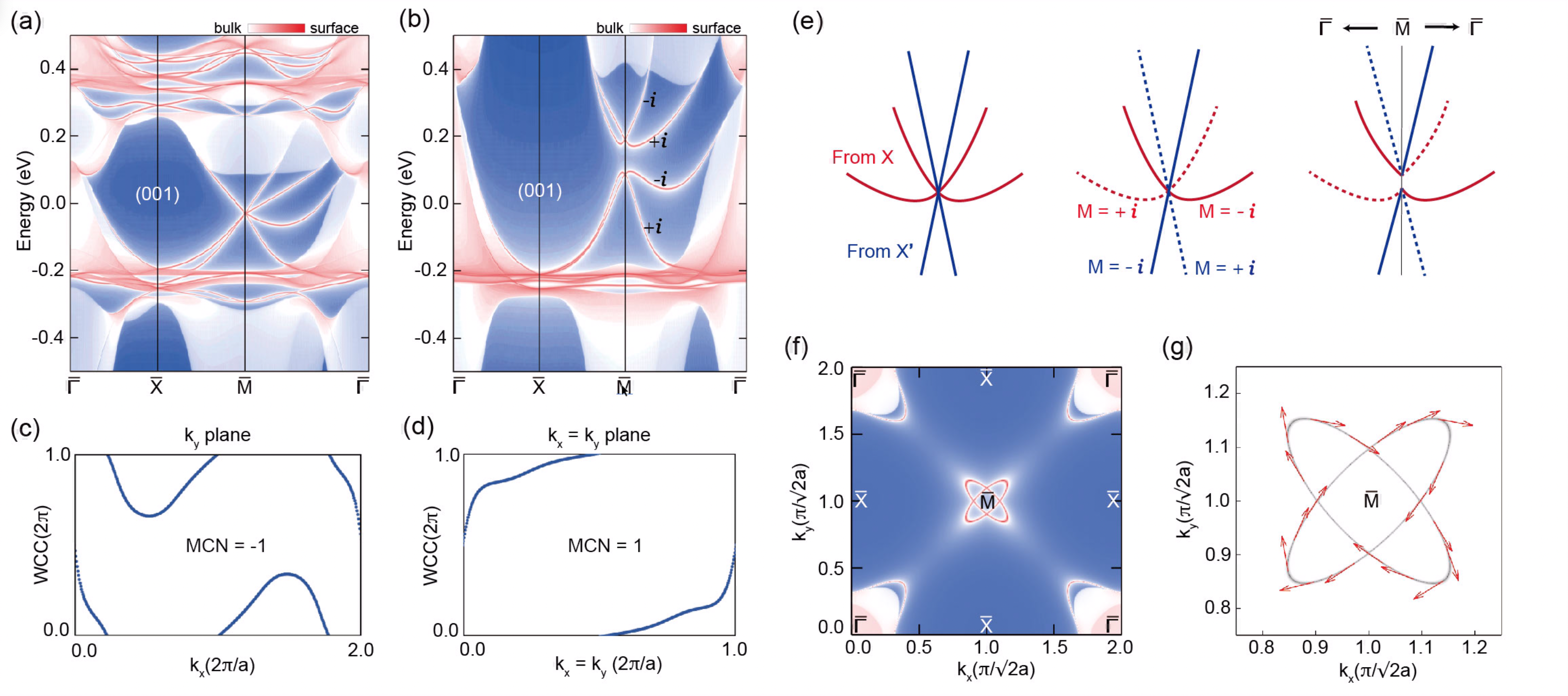}
\caption{(Color Online)
Semi-infinite TB slab calculations
for the (001) surface of $g$-SmS
with (a) normal SOC and (b) the enhanced SOC strength by ten times.
While the gap in the double Dirac cones in (a) is too tiny
to be identified,
the gap is clearly shown in (b) with the enhanced SOC strength.
Mirror eigenvalues along $\bar{\Gamma}-\bar{M}$ are denoted in (b).
(c),(d) The evolutions of the Wannier charge centers (WCCs),
respectively, for $k_{y} = 0$ and $k_{x} = k_{y}$
mirror-symmetry planes.
Two distinct MCNs, $(C_{0}, C_{d})=(-1, +1)$,
are obtained from the Wilson-loop calculations.
(e) Schematic diagram for the gap-opening mechanism
in the double Dirac cones along $\bar{\Gamma}-\bar{M}-\bar{\Gamma}$
(see Fig.~\ref{BZ}).
The bands having the same mirror eigenvalue (M)
hybridize with each other to produce the energy gap.
(f),(g) The FS, and its spin texture around $\bar{M}$.
The spin texture shows the Rashba-type spin polarization.
For the comparison to experiment, one needs to take into account
the scale factor of $Z \approx$ 0.1 in the $y$-axis DFT energies in
(a) and (b) due to the band renormalization effect.
}
\label{SmS1}
\end{figure*}

Another candidate for a topological Kondo material was
proposed theoretically
in a Sm mixed-valent/Kondo system, golden phase of SmS ($g$-SmS).
But, unlike SmB$_{6}$ that has a simple-cubic structure,
$g$-SmS crystallizes in rock-salt-type face-centered cubic (fcc) structure
and has a semi-metallic electronic structure.
The difference in crystal symmetry gives rise to a richer
or more intricate topological structure.
Due to the odd number of band inversions in the bulk band structure,
$g$-SmS was suggested to be a topological compensated semimetal \cite{Li14}.
However, (001) surface states exhibit double Dirac cones
with a tiny gap instead of gapless Dirac cones \cite{Kang15}.
This result is in contrast with that of
Kasinathan \emph{et al.}~\cite{Kasinathan15},
who reported the existence of gapless Dirac cones
on the (001) surface of the isoelectronic compound SmO.
Thus whether a gap exists in the double Dirac cones
on the (001) surface of $g$-SmS is still unclear,
which is an important issue to be clarified
in connection with its topological nature.

The double Dirac cones appear when two single-Dirac cones,
which are induced from band inversion
at non-equivalent bulk $k$-points,
are projected onto one $k$-point in the surface Brillouin zone (BZ)
(see Fig.~\ref{BZ}).
The double Dirac cones were detected in ARPES measurements
for CeBi \cite{Kuroda18,Li18}
and also for topological-crystalline insulators (TCIs) \cite{Fu11}
of SnTe \cite{Hsieh12,Tanaka12,Xu12} and SnSe \cite{Dziawa12,Okada13},
having the same rock-salt structure as $g$-SmS.
Note that double Dirac cones observed in the Sn-chalcogenides
have basically different topology from those in CeBi.
The Sn-chalcogenides have the gapless double Dirac cones,
while CeBi has the gapped double Dirac cones.
Therefore, it is imperative to identify
the topological nature of the double Dirac cones realized in $g$-SmS.

In this work, we have investigated the surface
states of $g$-SmS, which is one of the representative
Sm compounds exhibiting mixed-valent Kondo properties,
employing the density functional theory (DFT)
and the Wilson-loop calculations.
The strong correlation effect
of $4f$ electrons in Kondo systems can be captured by
renormalizing DFT $4f$ bands with a reasonable scale factor.
Indeed we  have shown before that the DFT band structures
near the Fermi level ({\EF}) have similar shape
to those obtained by the dynamical mean-field theory (DMFT)
at low temperature after rescaling the DFT band with the
DMFT renormalization factors \cite{Denlinger13,Kang15,Kim14,Kang15JPSJ}.
With this rescaling in the bulk, the slab calculations properly describe
the surface states \cite{Zfac}.

Here we have demonstrated for $g$-SmS that
(i) it has nontrivial mirror Chern numbers (MCNs),
(ii) the (001) surface has the gapped double Dirac cones
seemingly of Rashba-type,
(iii) the (111) surface has a clear single Dirac cone
in the gap region,
confirming that $g$-SmS is indeed a topological Kondo system,
and (iv) the (110) surface has the TCI-type double
Dirac cones as well as
the intriguing topological-insulator (TI)-type single Dirac cone.

\begin{table}[b]
\caption{
Products of parity eigenvalues of the occupied states
at the time-reversal invariant momentum (TRIM) points
of the bulk BZ of fcc $g$-SmS.
It indicates nontrivial $Z_2$ topology.
}
\begin{center}
\begin{ruledtabular}
\begin{tabular}{ccccc}
& $\Gamma$ & 3~$X$ & 4~$L$ &  Z$_2$ \\
\hline
$g$-SmS & + & $-$ & + &$1$
\end{tabular}
\end{ruledtabular}
\end{center}
\label{parity}
\end{table}

\emph{Method}. For the DFT calculations, we have used the
projector-augmented wave band method \cite{PAW}, implemented in VASP \cite{Vasp}.
We have employed the generalized-gradient approximation \cite{GGA}
for the exchange-correlation functional.
A lattice constant of $a = 5.6~\AA$ was used for $g$-SmS.
To investigate surface electronic structures,
we have constructed the tight-binding (TB) Hamiltonian
from DFT results,
using the Wannier interpolation scheme
implemented in WANNIER90 code \cite{Wannier90},
and then performed semi-infinite TB slab calculations using
the Green function method \cite{Green} implemented
in WannierTools \cite{Wtool}.
We have double-checked the surface band structures
by performing the DFT slab calculations with
both the WIEN2K \cite{Wien2k} and the VASP \cite{Vasp} codes.
The topological nature of surface states is analyzed in terms of
the mirror eigenvalues and MCNs
\cite{Ye13,Legner14,Legner15,Baruselli15,Baruselli16},
which are obtained
from the Wilson-loop calculations \cite{Yu11,Soluyanov11}
based on the model-independent \emph{ab initio} calculations.

\emph{(001) surface}.
Since $g$-SmS shows
Sm $4f$-$5d$ band inversion at $X$ of the bulk BZ
\cite{Li14,Kang15},
it provides a nontrivial Z$_{2}$ number, as shown in Table~\ref{parity}.
On the (001) surface of $g$-SmS,
one $X$ point is projected onto  $\bar{\Gamma}$,
while two non-equivalent bulk $X$ points
($X$ and $X^{\prime}$ in Fig.~\ref{BZ})
are projected onto the $\bar{M}$ point of the surface BZ,
and so the single and double Dirac cones are expected to
be realized, respectively, at $\bar{\Gamma}$ and $\bar{M}$.

Figure \ref{SmS1} shows the (001) surface
band structures obtained from semi-infinite TB slab calculations.
The single Dirac cone at $\bar{\Gamma}$
is hardly detectable in Fig.~\ref{SmS1}(a) due to
an overlap with the bulk band structures.
On the other hand, the double Dirac cones are noticeable at $\bar{M}$,
which seem to be gapless as claimed by
Kasinathan \emph{et al.}~\cite{Kasinathan15} for SmO.
However, a tiny band gap actually exists.
The gap opening is clearly identified
for the calculation with ten-times enhanced
spin-orbit coupling (SOC) strength in Fig.~\ref{SmS1}(b) \cite{10xsoc},
which manifests the gapped double Dirac cones of Rashba-type
in agreement with a previous report \cite{Kang15}.
As shown in Fig.~\ref{BZ}, on the (001) surface,
there are two mirror-symmetry lines along $\bar{\Gamma}-\bar{X}$
and $\bar{\Gamma}-\bar{M}$.
Therefore, the gap opening along $\bar{X}-\bar{M}$ is obvious
because there is no mirror symmetry to protect the band crossings.
But, further analysis is required
for the surface states along $\bar{\Gamma}-\bar{M}$.

To explore the origin of the gap opening along $\bar{\Gamma}-\bar{M}$,
we have calculated MCNs and the mirror eigenvalues of the surface states.
In the bulk BZ, there are two independent mirror-symmetry planes:
the $k_{y} = 0$ and $k_{x} = k_{y}$ planes \cite{mirror}.
Note that two mirror operators with respect to the $k_{y} = 0$
and $k_{y} = \pi$ planes are symmetrically equivalent.
Therefore $g$-SmS has two independent MCNs,
$C_{0}(\equiv C_{k_y = 0}^{+}$)
and $C_{d}(\equiv C_{k_x = k_y}^{+}$),
where the `$+$' sign refers to
the mirror eigenvalue of $+i$
for the corresponding mirror-symmetry plane.

\begin{figure}[t]
\includegraphics[width=8.5 cm]{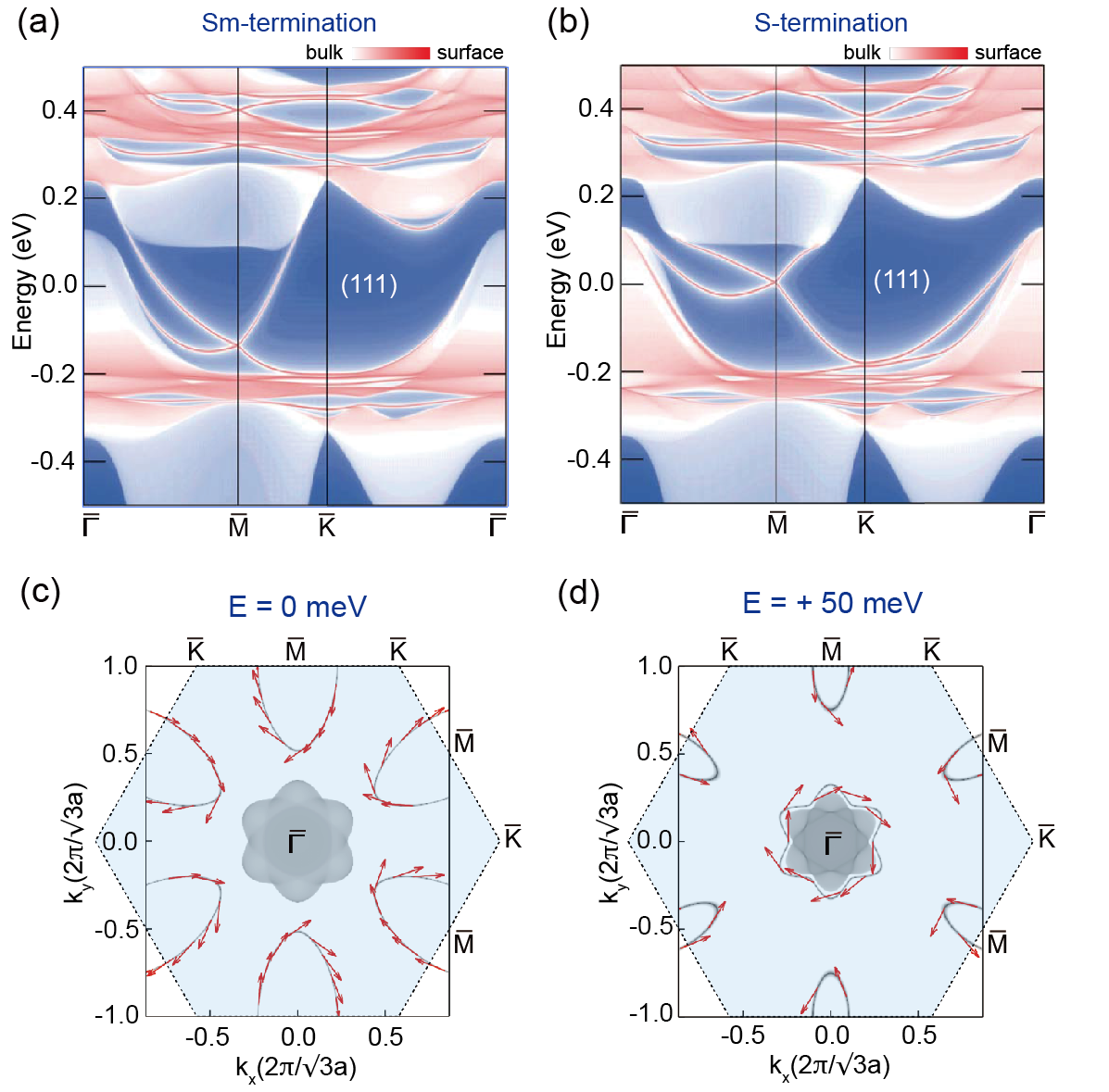}
\caption{(Color Online)
Surface electronic structures from the
semi-infinite TB slab calculations
for the (111) surface of $g$-SmS
with (a) Sm- and (b) S-termination.
Topological Dirac-cone surface states are clearly shown
at $\bar{M}$ in both (a) and (b).
(c),(d) The FS and the energy contour at $E= 50$ meV,
respectively, for the Sm-terminated case.
Their spin textures with spin helicities of Rashba-type
are also provided.
}
\label{SmS2}
\end{figure}

Figures \ref{SmS1}(c) and \ref{SmS1}(d) show the evolutions of the
Wannier charge centers (WCCs)
for the $k_{y} = 0$ and $k_{x} = k_{y}$ mirror planes.
The corresponding MCNs are obtained to be
$C_{0}=-1$ and $C_{d}=1$,
which implies the existence
of at least one gapless Dirac cone along
both $\bar{M}-\bar{\Gamma}-\bar{M}$
and $\bar{X}-\bar{\Gamma}-\bar{X}$ lines
on the (001) surface (see Fig.~\ref{BZ}).
Accordingly, the MCN of $C_{0}=-1$ is suggestive of
ruling out the existence of additional gapless Dirac cones
along $\bar{\Gamma}-\bar{M}$
besides the gapless single Dirac cone buried at $\bar{\Gamma}$.
Indeed, mirror eigenvalue analysis provides
a more direct explanation on the gapping.
Depicted schematically in Fig.~\ref{SmS1}(e)
are the double Dirac cones along $\bar{\Gamma}-\bar{M}-\bar{\Gamma}$,
which are composed of two single Dirac cones
arising from band inversions at
two non-equivalent bulk $X$ points, $X$ (red) and $X^{\prime}$ (blue).
Two bands of each single Dirac cone have opposite mirror eigenvalues:
$+i$ for the dotted and $-i$ for the solid line in Fig.~\ref{SmS1}(e).
Then the crossing Dirac-cone bands with the same mirror eigenvalues
hybridize with each other
to produce the hybridization gap around $\bar{M}$,
resulting in the gapped double Dirac cones.
Namely, the double Dirac cones at $\bar{M}$
have neither the TI nor the TCI character \cite{Fu11}.
As mentioned earlier, this kind of double Dirac-cone feature
around $\bar{M}$ was detected in ARPES for CeBi \cite{Kuroda18,Li18}.
Note, however, that the double Dirac cones detected in CeBi
appear due to a $p$-$d$ (not $f$-$d$) band inversion.

Figure \ref{SmS1}(f) presents the Fermi surface (FS)
of the (001) surface band structure with normal SOC strength.
The crossing oval-shaped FS is apparent around $\bar{M}$,
which arises from the double Dirac cones at $\bar{M}$
in Fig.~\ref{SmS1}(a).
On the other hand, the FS around $\bar{\Gamma}$
is derived from both bulk and surface band structures.
The spin texture of the FS around $\bar{M}$ is provided
in Fig.~\ref{SmS1}(g).
The spin-helical structure around each ellipse is evident,
reflecting that the gapped double Dirac cones have
the spin texture of Rashba-type (not Dresselhaus-type)
\cite{Legner14,Legner15,Baruselli15,Baruselli16}.

\begin{figure*}[t]
\includegraphics[width=17 cm]{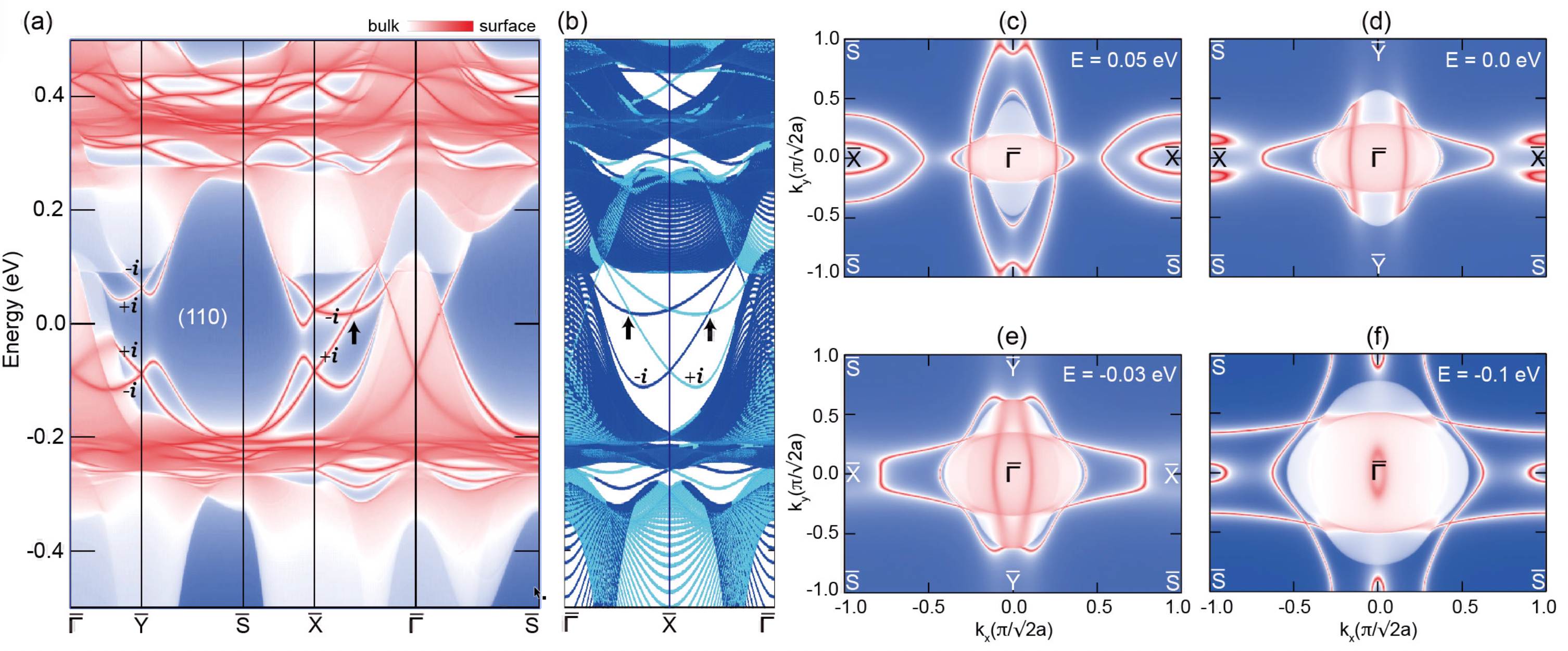}
\caption{(Color Online)
(a) Semi-infinite TB slab calculations
for the (110) surface of $g$-SmS.
The double Dirac cones of TCI-type are clearly manifested around $\bar{X}$.
A single Dirac-cone surface state is also seen at $\bar{\Gamma}$,
even though it is buried inside the bulk-projected bands.
(b) Mirror eigenvalues of the double Dirac cones
along $\bar{\Gamma}-\bar{X}-\bar{\Gamma}$.
Mirror eigenvalues of $+i$ and $-i$ are presented in
aqua and navy-blue colors, respectively.
(c)-(f) The FS and energy contours on the (110) surface.
}
\label{SmS110}
\end{figure*}

\emph{(111) surface}. In order to ascertain the topological nature
of $g$-SmS more evidently,
we have investigated the (111) surface states.
As shown in Fig.~\ref{BZ},
each non-equivalent bulk $X$ point is projected onto
a different $\bar{M}$ point of surface BZ on the (111) surface.
Hence, the gapless single Dirac cone
could be realized at each $\bar{M}$ point.
Note that the (111) surface has two kinds of terminations:
Sm- and S-terminations.
As shown in Figs.~\ref{SmS2}(a) and \ref{SmS2}(b),
both terminations indeed possess the gapless
single Dirac cone at $\bar{M}$ in the gap region.
Figures~\ref{SmS2}(c) and \ref{SmS2}(d) show the FS
and the energy contour at $E=50$ meV, respectively,
for the Sm-terminated case.
They clearly reveal the helical spin textures of Rashba-type,
which originate from the single Dirac-cone surface states.
These features provide unambiguous evidence
of the topological nature in semimetallic $g$-SmS.

\emph{(110) surface}.
As in the case of the (001) surface,
two non-equivalent bulk $X$ points ($X$ and $X^{\prime}$)
are projected onto $\bar{X}$ of the (110) surface BZ (see Fig.~\ref{BZ}).
So the (110) surface also has double Dirac cones at $\bar{X}$.
It is thus worthwhile to check whether these double
Dirac cones at $\bar{X}$ would produce
the TCI-type or
the Rashba-type surface states as in the (001) surface.
As shown in Fig.~\ref{SmS110}(a),
the double Dirac cones at $\bar{X}$ manifest a hallmark
of TCI-type surface states with a Dirac point
off the TRIM points (marked by an arrow).
They are gapped along $\bar{S}-\bar{X}$
because $\bar{S}-\bar{X}$ is not a mirror-symmetry line,
while, along the mirror-symmetry line $\bar{X}-\bar{\Gamma}$,
they show the prominent Dirac point in the gap region.
To ensure the band crossing in-between $\bar{X}-\bar{\Gamma}$,
we have analyzed their mirror eigenvalues.
It is shown in Fig.~\ref{SmS110}(b) that the crossing surface bands
have opposite mirror eigenvalues, $+i$ and $-i$,
so that the band crossings are protected by the mirror symmetry,
which is distinct from the case of the (001) surface.
Note that, besides the double Dirac cones at $\bar{X}$,
the (110) surface exhibits a single Dirac cone of TI-type
at $\bar{\Gamma}$ in Fig.~\ref{SmS110}(a).
Its Dirac point is clearly manifested at $\bar{\Gamma}$,
even though it is buried inside the bulk-projected bands
at $E \approx -0.1$ eV.

As shown in Fig.~\ref{SmS110}(c)-(f),
the energy contours around $\bar{X}$ display the Lifshitz-like transition
as a function of binding energies,
which is another manifestation
of the TCI-type double Dirac-cone surface states \cite{Hsieh12}.
However, the energy contours around $\bar{\Gamma}$ look
more complicated than a single Dirac cone.
This intricate features are expected to come from the
TI-type single Dirac-cone surface state distorted along
the other mirror-symmetry line, $\bar{\Gamma}-\bar{Y}$.
As shown in Fig.~\ref{SmS110}(a),
two neighboring surface states along $\bar{\Gamma}-\bar{Y}$,
one of which corresponds to the Dirac-cone state
and the other to a trivial state,
have the same mirror eigenvalues, $+i$,
and so they are hybridized to be gapped.
Hence, most interestingly, the (110) surface has both
the TCI and the intricate TI nature.
It is noteworthy that this kind of TI/TCI feature was also
proposed in the (110) surface of SmB$_{6}$
\cite{Ye13,Legner14,Legner15,Baruselli15,Baruselli16}.
But one should keep in mind that they have different
crystal structures, fcc $g$-SmS vs simple-cubic SmB$_{6}$,
which induce different mirror symmetries in the two materials.


Finally, we would like to comment on the experimental verification
of the topological nature of $g$-SmS.
Note that $g$-SmS is a phase under pressure \cite{Kang15}.
Therefore, it is not easy to probe its topological fingerprints
by employing conventional ARPES because it is difficult to
apply external pressure in ARPES.
Instead of applying external pressure,
one may utilize chemical pressure or strain
to simulate the $g$-SmS phase with a reduced volume.
For example, Y-substituted SmS (Sm$_{1-x}$Y$_x$S) or SmS-film
grown on an iso-structural sulphide having a smaller volume
can be used.
In fact, ARPES measurements have been performed
on metallic Sm$_{1-x}$Y$_x$S \cite{Imura13,Kaneko14},
but no surface state has been observed yet.

\emph{Conclusion}.
We have demonstrated that $g$-SmS has the gapless single Dirac cone
in the gap region of its (111) surface BZ, which provides unambiguous evidence
of the topological Kondo nature in mixed-valent $g$-SmS.
The double Dirac cones realized in the (001) and (110) surfaces of $g$-SmS
result in the Rashba-type and TCI-type surface states, respectively, which are
elaborated by the mirror eigenvalues and MCNs, obtained by \emph{ab initio}
band structure calculations.
It is thus worth challenging experimentalists to identify,
via high-resolution ARPES,
the gapped double Dirac cones of Rashba-type surface states,
the gapless TI-type Dirac cone,
and the TCI-type double Dirac cones,
respectively, for the (001), (111), and (110) surfaces of $g$-SmS
or Sm$_{1-x}$Y$_{x}$S.


\emph{Acknowledgments}.
Chang-Jong Kang and Dong-Choon Ryu contributed equally
to this work.
This work was supported by the National Research Foundation
(NRF) Korea (Grants No. 2016R1D1A1B02008461,
No. 2017R1A2B4005175, and No. 2019R1A2C1004929),
Max-Plank POSTECH/KOREA Research Initiative
(Grant No. 2016K1A4A4A01922028), the POSTECH BSRI Grant,
and KISTI supercomputing center (Grant No. KSC-2017-C3-0057).
The ALS is supported by U.S. DOE  under Contract No.
DE-AC02-05CH11231.
C.-J.K. and G.K. were supported by the National Science Foundation Grant DMR-1733071.




\begin{thebibliography}{99}
\bibitem{Zfac}
    For $g$-SmS, a proper value of $Z$ is known to be around 0.1
    (see Ref.~\cite{Kang15}).
\bibitem{10xsoc} We used ten-times larger SOC in Fig.~\ref{SmS1}(b)
	to enhance the visibility of an existing tiny gap.
 	It could, however, be a better representation of reality
	by shifting up the SOC-split $j=7/2$ states so as
	to remove incorrect interaction with surface
	states near {\EF}.
\bibitem{mirror} Here we conisdered the (001) surface instead of the (100) surface.
\end{thebibliography}
\end{document}